\documentclass[preprint,11pt]{elsarticle}

\journal{Information and Software Technology}

\usepackage{todonotes}
\usepackage{amsmath,amssymb,amsfonts}
\usepackage{algorithmic}
\usepackage{graphicx}
\usepackage{textcomp}
\usepackage{xcolor}
\usepackage{multirow}
\usepackage{tabularx}
\usepackage{array}
\usepackage{flushend}
\usepackage{balance}
\usepackage{lscape}
\newcolumntype{?}{!{\vrule width 1pt}}
\usepackage{float}
\newcolumntype{C}[1]{>{\centering\arraybackslash}p{#1}}
\usepackage{subcaption}
\def\BibTeX{{\rm B\kern-.05em{\sc i\kern-.025em b}\kern-.08em
    T\kern-.1667em\lower.7ex\hbox{E}\kern-.125emX}}
\graphicspath{ {./images/} }
\usepackage{tikz}
\tikzstyle{mybox} = [draw=black, very thick, rectangle, rounded corners, inner ysep=5pt, inner xsep=5pt]

\begin{document}

\begin{frontmatter}

\title{Does Code Quality Affect Pull Request Acceptance? An empirical study}

\author {Valentina Lenarduzzi}
\ead{valentina.lenarduzzi@tuni.fi}

\author {Vili Nikkola}
\ead{vili.nikkola@tuni.fi}

\author {Nyyti Saarim\"{a}ki}
\ead{nyyti.saarimaki@tuni.fi}

\author {Davide Taibi}
\ead{davide.taibi@tuni.fi}

\address {Tampere University, Tampere (Finland)}

\begin{abstract}
\textit{Background.} Pull requests are a common practice  for contributing and reviewing contributions, and are employed both in  open-source and industrial contexts.
One of the main goals of code reviews is to find defects in the code, allowing  project maintainers to easily integrate external contributions into a project and discuss the code contributions. 

\noindent\textit{Objective}. The goal of this paper is to understand whether code quality is actually considered when pull requests are accepted. Specifically, we aim at understanding whether code quality issues such as code smells, antipatterns, and coding style violations in the pull request code affect the chance of its acceptance when reviewed by a maintainer of the project.

\noindent\textit{Method}. We conducted a case study among 28 Java open-source projects, analyzing the presence of 4.7 M code quality issues in 36 K pull requests. We analyzed further correlations by applying Logistic Regression and seven machine learning techniques (Decision Tree,  Random Forest, Extremely Randomized Trees, AdaBoost, Gradient Boosting, XGBoost).

\noindent\textit{Results}. Unexpectedly, code quality turned out not to affect the acceptance of a pull request at all. As suggested by other works, other factors such as the reputation of the maintainer and the importance of the feature delivered might be more important than code quality in terms of pull request acceptance. 

\noindent\textit{Conclusions}. Researchers already investigated the influence of the developers' reputation and the pull request acceptance. This is the first work investigating if quality of the code in pull requests affects the acceptance of the pull request or not. We recommend that researchers further investigate this topic to understand if different measures or different tools could provide some useful measures.  
\end{abstract}


\end{frontmatter}


\section{Introduction}
\label{Introduction}
Different code review techniques have been proposed in the past and widely adopted by open-source and commercial projects. Code reviews involve the manual inspection of the code by different developers and help companies to reduce the number of defects and improve the quality of software~\cite{Ackerman1984}\cite{Ackerman1989}.




Nowadays, code reviews are generally no longer conducted as they were in the past, when developers organized review meetings to inspect the code line by line~\cite{Fagan1976}.

Industry and researchers agree that code inspection helps to reduce the number of defects, but that in some cases, the effort required to perform code inspections hinders their adoption in practice~\cite{Shull2008}.
However, the born of new tools and 
has enabled companies to adopt different code review practices. In particular, several companies, including Facebook~\cite{Feitelson2013}, Google~\cite{Potvin2016}, and Microsoft~\cite{Bacchelli2013}, perform code reviews by means of tools such as   Gerrit\footnote{https://www.gerritcodereview.com} or by means of the pull request mechanism provided by Git\footnote{https://help.github.com/en/articles/about-pull-requests}~\cite{Rigby}.

 In the context of this paper, we focus on pull requests. Pull requests provide developers a convenient way of contributing to projects, and many popular projects, including both open-source and commercial ones, are using pull requests as a way of reviewing the contributions of different developers.




Researchers have focused their attention on pull request mechanisms, investigating different aspects, including the review  process~\cite{Gousios2014}, \cite{Gousios2015} and \cite{Veen2015}, the influence of code reviews on continuous integration builds~\cite{Zampetti2017}, how pull requests are assigned to different reviewers~\cite{Yu2014}, and in which conditions they are accepted process~\cite{Gousios2014},\cite{Rahman2014},\cite{Soares2015},\cite{Kononenko2018}. 
Only a few works have investigated whether developers consider quality aspects in order to accept pull requests~\cite{Gousios2014},\cite{Gousios2015}. 
Different works report that the reputation of the developer who submitted the pull request is one of the most important acceptance factors~\cite{Gousios2015},\cite{Calefato2017}. 


However, to the best of our knowledge, no studies have investigated whether the quality of the code submitted in a pull request has an impact on the acceptance of this pull request. As code reviews are a fundamental aspect of pull requests, we strongly expect that pull requests containing low-quality code should generally not be accepted. 

In order to understand whether code quality is one of the acceptance drivers of pull requests, we designed and conducted a case study involving 28 well-known Java projects to analyze the quality of more than 36K pull requests. 
We analyzed the quality of pull requests using PMD\footnote{https://pmd.github.io}, one of the four tools used most frequently for software analysis~\cite{LenarduzziSEDA2019}, \cite{Beller2016}.
PMD evaluates the code quality against a standard rule set available for the major languages, allowing the detection of different quality aspects generally considered harmful, including code smells~\cite{Fowler1999} such as   ''long methods'', ''large class'', ''duplicated code''; anti-patterns~\cite{BrownAntipatterns} such as ''high coupling''; design issues such as ''god class''~\cite{LanzaMarinescu}; and various coding style violations\footnote{https://pmd.github.io/latest/pmd\_rules\_java.html}. Whenever a rule is violated, PMD raises an issue that is counted as part of the Technical Debt~\cite{Cunningham1992}. In the remainder of this paper, we will refer to all the issues raised by PMD as ''TD items'' (Technical Debt items). 

Previous work confirmed that the presence of several code smells and anti-patterns, including those collected by PMD, significantly increases the risk of faults on the one hand and maintenance effort on the other hand~\cite{Khomh2009}, \cite{Olbrich2009}, \cite{DAmbros2010}, \cite{Fontana2011}.


Unexpectedly, our results show that the presence of TD items of all types does not influence the acceptance or rejection of a pull request at all. 
Based on this statement, we analyzed all the data not only using basic statistical techniques, but also applying seven machine learning algorithms (Logistic Regression, Decision Tree,  Random Forest, Extremely Randomized Trees, AdaBoost, Gradient Boosting, XGBoost), analyzing 36,986 pull requests and over 4.6 million TD items present in the pull requests.

\textbf{Structure of the paper}. Section~\ref{Background} describes the basic concepts underlying this work, while Section~\ref{RW} presents some related work done by researchers in recent years.  In  Section~\ref{CS}, we describe the design of our case study, defining the research questions, metrics, and hypotheses, and describing the study context, including the data collection and data analysis protocol. In Section~\ref{Results}, we present the achieved results and discuss them in Section~\ref{Discussion}. Section~\ref{Threats} identifies the threats to the validity of our study, and in  Section~\ref{Conclusion}, we draw conclusions and give an outlook on possible future work. 

\section{Background}
\label{Background}
In this Section, we will first introduce code quality aspects and PMD, the tool we used to analyze the code quality of the pull requests. Then we will describe the pull request mechanism and finally provide a brief introduction and motivation for the usage of the machine learning techniques we applied.

\subsection{Code Quality and PMD}
\label{PMD}

Different tools on the market can be used to evaluate code quality. PMD is one of the most frequently used static code analysis tools for Java on the market, along with
Checkstyle, Findbugs, 
 and SonarQube~\cite{LenarduzziSEDA2019}. 

PMD is an open-source tool that aims to identify issues that can lead to technical debt accumulating during development. The specified source files are analyzed and the code is checked with the help of predefined rule sets. PMD  provides a standard rule set for major languages, which the user can customize if needed. The default Java rule set encompasses all available Java rules in the PMD project and is used throughout this study.

Issues found by PMD have five priority values (P). Rule priority guidelines for default and custom-made rules can be found in the PMD project documentation \textsuperscript{4}.
\begin{enumerate}
\label{list:priorities}
\item [P1]Change absolutely required. Behavior is critically broken/buggy.
\item [P2]Change highly recommended. Behavior is quite likely to be broken/buggy.
\item [P3]Change recommended. Behavior is confusing, perhaps buggy, and/or against standards/best practices.
\item [P4]Change optional. Behavior is not likely to be buggy, but more just flies in the face of standards/style/good taste.
\item [P5]Change highly optional. Nice to have, such as a consistent naming policy for package/class/fields…
\end{enumerate}

These priorities are used in this study to help determine whether more severe issues affect the rate of acceptance in pull requests.

PMD is the only tool that does not require compiling the code to be analyzed. This is why, as the aim of our work was to analyze only the code of pull requests instead of the whole project code, we decided to adopt it. PMD defines more than 300 rules for Java, classified in eight categories (coding style, design, error prone, documentation, multithreading, performance, security). Several rules have also been confirmed harmful by different empirical studies. 
In Table I
we highlight a subset of rules and the related  empirical studies that confirmed their harmfulness. The complete set of rules is available on the PMD official documentation\textsuperscript{4}.

\begin{table}
\label{tab:pmd-rules}
\caption {Example of PMD rules and their related harmfulness} 
\begin{center}
\footnotesize
\begin{tabular}{p{3.8cm}|p{4.5cm}|p{3.7cm}}
\hline
\textbf{PMD Rule} & \textbf{Defined By} & \textbf{Impacted Characteristic}  \\ \hline

Avoid Using Hard-Coded IP	&	Brown et al~\cite{Brown}	&	Maintainability~\cite{Brown}	\\ \hline

Loose Coupling	&	Chidamber and Kemerer~\cite{Chidamber1994}	&	Maintainability~\cite{AlDallal2018}	\\ \hline

Base Class Should be Abstract	&	Brown et al~\cite{Brown}	&	Maintainability~\cite{Khomh2009}	\\ \hline
Coupling Between Objects	&	Chidamber and Kemerer~\cite{Chidamber1994}	&	Maintainability~\cite{AlDallal2018}	\\ \hline
Cyclomatic Complexity	&	Mc Cabe~\cite{McCabe1976}	&	Maintainability~\cite{AlDallal2018}	\\ \hline
Data Class	&	Fowler~\cite{Fowler1999}	&	Maintainability~\cite{Li2007}, \\ 
&& Faultiness \cite{Sjoberg2013}, \cite{yamashita2014assessing}	\\ \hline
Excessive Class Length	&	Fowler (Large Class)~\cite{Fowler1999}	&	Change Proneness~\cite{Palomba2018}, \cite{Kohm2009} 	\\ \hline
Excessive Method Length	&	Fowler (Large Method)~\cite{Fowler1999}	&	Change Proneness~\cite{Jaafar2016}, \cite{Kohm2009} Fault Proneness \cite{Palomba2018} 	\\ \hline
Excessive Parameter List	&	Fowler (Long Parameter List)~\cite{Fowler1999}	&	Change Proneness~\cite{Jaafar2016}	\\ \hline
God Class 	&	Marinescu and Lanza~\cite{LanzaMarinescu}	&	Change Pronenes \cite{Olbrich2010}, \cite{Schumacher2010}, \cite{Zazworka2010},  Comprehensibility \cite{DuBoisDVMT06}, \\
&& Faultiness \cite{Olbrich2010}\cite{Zazworka2010}	\\ \hline
Law of Demeter 	&	Fowler (Inappropriate Intimacy)~\cite{Fowler1999}	&	Change Proneness~\cite{Palomba2018} 	\\ \hline
Loose Package Coupling	&	Chidamber and Kemerer~\cite{Chidamber1994} &	Maintainability~\cite{AlDallal2018}	\\ \hline
Comment Size	&	Fowler (Comments) \cite{Fowler1999}	& Faultiness \cite{Aman2014}, \cite{Aman2012}	\\
\hline
\end{tabular}
\end{center}
\end{table}

\subsection{Git and Pull Requests}
\label{Git}
Git\footnote{https://git-scm.com/} is a distributed version control system that enables users to collaborate on a coding project by offering a robust set of features to track changes to the code. Features include “committing” a change to a local repository, “pushing” that piece of code to a remote server for others to see and use, “pulling” other developers’ change sets onto the user's workstation, and merging the changes into their own version of the code base. Changes can be organized into branches, which are used in conjunction with pull requests. Git provides the user a "diff" between two branches, which compares the branches and provides an easy method to analyze what kind of additions the pull request will bring to the project if accepted and merged into the master branch of the project.

Pull requests are a code reviewing mechanism that is compatible with Git and are provided by GitHub\footnote{https://github.com/}. The goal is for code changes to be reviewed before they are inserted into the mainline branch. A developer can take these changes and push them to a remote repository on GitHub. Before merging or rebasing a new feature in, project maintainers in GitHub can review, accept, or reject a change based on the diff of the “master” code branch and the branch of the incoming change. Reviewers can comment and vote on the change in the GitHub web user interface. If the pull request is approved, it can be included in the master branch. A rejected pull request can be abandoned by closing it or the creator can further refine it based on the comments given and submit it again for review.

\subsection{Machine Learning Techniques}
In this section, we will describe the  machine learning classifiers adopted in this work. We used eight different classifiers: a generalized linear model (Logistic Regression), a tree-based classifier (Decision Tree), and six ensemble classifiers (Bagging, Random Forest, ExtraTrees, AdaBoost, GradientBoost, and XGBoost). 


In the next sub-sections, we will briefly introduce the eight adopted classifiers and give the rationale for choosing them for this study.

\textit{Logistic Regression} \label{LR}~\cite{Cox1958} is one of the most frequently used algorithms in Machine Learning. In logistic regression, a collection of measurements (the counts of a particular issue) and their binary classification (pull request acceptance) can be turned into a function that outputs the probability of an input being classified as 1, or in our case, the probability of it being accepted.


\textit{Decision Tree}\label{DT}
~\cite{breiman1984classification} is a model that takes learning data and constructs a tree-like graph of decisions that can be used to classify new input. The learning data is split into subsets based on how the split from the chosen variable improves the accuracy of the tree at the time. The decisions connecting the subsets of data form a flowchart-like structure that the model can use to tell the user how it would classify the input and how certain the prediction is perceived to be.

We considered two methods for determining how to split the learning data: GINI impurity and information gain. GINI tells the probability of an incorrect classification of a random element from the subset that has been assigned a random class within the subset. Information gain tells how much more accuracy a new decision node would add to the tree if chosen. GINI was chosen because of its popularity and its resource efficiency.

Decision Tree as a classifier was chosen because it is easy to implement and human-readable; also, decision trees can handle noisy data well because subsets without significance can be ignored by the algorithm that builds the tree. The classifier can be susceptible to overfitting, where the model becomes too specific to the data used to train it and provides poor results when used with new input data. Overfitting can become a problem when trying to apply the model to a mode-generalized dataset.

\textit{Random Forest}
\label{RF}~\cite{Breiman2001} is an ensemble classifier, which tries to reduce the risk of overfitting a decision tree by constructing a collection of decision trees from random subsets in the data. The resulting collection of decision trees is smaller in depth, has a reduced degree of correlation between the subset's attributes, and thus has a lower risk of overfitting.

When given input data to label, the model utilizes all the generated trees, feeds the input data into all of them, and uses the average of the individual labels of the trees as the final label given to the input.

\textit{Extremely Randomized Trees}
\label{ET}
~\cite{Geurts2006} builds upon the Random Forest introduced above by taking the same principle of splitting the data into random subsets and building a collection of decision trees from these. In order to further randomize the decision trees, the attributes by which the splitting of the subsets is done are also randomized, resulting in a more computationally efficient model than Random Forest while still alleviating the negative effects of overfitting.

\textit{Bagging}
~\label{BG}\cite{Breiman1996} is an ensemble classification technique that tries to reduce the effects of overfitting a model by creating multiple smaller training sets from the initial set; in our study, it creates multiple decision trees from these sets. The sets are created by sampling the initial set uniformly and with replacements, which means that individual data points can appear in multiple training sets. The resulting trees can be used in labeling new input through a voting process by the trees.

\textit{AdaBoost}
~\label{AB}\cite{FREUND1997119} is a classifier based on the concept of boosting. The implementation of the algorithm in this study uses a collection of decision trees, but new trees are created with the intent of correctly labeling instances of data that were misclassified by previous trees. For each round of training, a weight is assigned to each sample in the data. After the round, all misclassified samples are given higher priority in the subsequent rounds. When the number of trees reaches a predetermined limit or the accuracy cannot be improved further, the model is finished. When predicting the label of a new sample with the finished model, the final label is calculated from the weighted decisions of all the constructed trees. As Adaboost is based on  decision trees, it can be resistant to overfitting and be more useful with generalized data. However, Adaboost is susceptible to noise data and outliers.

\textit{Gradient Boost}
~\label{GB}\cite{friedman2001} is similar to the other boosting methods. It uses a collection of weaker classifiers, which are created sequentially according to an algorithm. In the case of Gradient Boost as used in this study, the determining factor in building the new decision trees is the use of a loss function. The algorithm tries to minimize the loss function and, similarly to Adaboost, stops when the model has been fully optimized or the number of trees reaches the predetermined limit.

\textit{XGBoost}
\label{XGB}
~\cite{Chen:2016:XST:2939672.2939785} is a scalable implementation of Gradient Boost. The use of XGBoost can provide performance improvements in constructing a model, which might be an important factor when analyzing a large set of data.

\section{Related Work}
\label{RW}
In this Section, we report on the most relevant works on pull requests. 
\subsection{Pull Request Process}
Pull requests have been studied from different points of view, such as pull-based development~\cite{Gousios2014}, \cite{Gousios2015} and \cite{Veen2015}, usage of real online resources~\cite{Zampetti2017}, pull requests reviewer assignment~\cite{Yu2014}, and acceptance process~\cite{Gousios2014}, \cite{Rahman2014}, \cite{Soares2015}, \cite{Kononenko2018}. 
Another issue regarding pull requests that have been investigated is latency. Yu et al.~\cite{Yu2015} define latency as a complex issue related to many independent variables such as the number of comments and the size of a pull request.

Zampetti et al.~\cite{Zampetti2017} investigated how, why, and when developers refer to online resources in their pull requests.
They focused on the context and real usage of online resources and how these resources have evolved during time. 
Moreover, they investigated the browsing purpose of online resources in pull request systems. 
Instead of investigating commit messages, they evaluated only the pull request descriptions, since generally the documentation of a change aims at reviewing and possibly accepting the pull request~\cite{Gousios2014}. 

Yu et al.~\cite{Yu2014} worked on pull requests reviewer assignment in order to provide an automatic organization in GitHub that leads to an effort waste. They proposed a reviewer recommender, who should predict highly relevant reviewers of incoming pull requests based on the textual semantics of each pull request and the social relations of the developers. They found several factors that influence pull requests latency such as size, project age, and team size.

This approach reached a precision rate of 74\% for top-1 recommendations, and a recall rate of 71\% for top-10 recommendations. However, the authors did not consider the aspect of code quality. The results are confirmed also by~\cite{Soares2015}. 

Recent studies investigated the factors that influence the acceptance and rejection of a pull request. 

There is no difference in treatment of pull-requests coming from the core team and from the community. Generally merging decision is postponed based on technical factors~\cite{Hellendoorn2015},\cite{Rigby2011}.
Generally, pull requests that passed the build phase are generally merged more frequently~\cite{Zampetti2019} 

Integrators decide to accept a contribution after analysing source code quality, code style, documentation, granularity, and adherence to project conventions~\cite{Gousios2014}.
Pull request's programming language had a significant influence on acceptance~\cite{Rahman2014}. Higher acceptance was mostly found for Scala, C, C\#, and R programming languages.
Factors regarding developers are related to acceptance process, such as the number and experience level of developers~\cite{Rahman2016}, and the developers reputation who submitted the pull request~\cite{Calefato2017}. 
Moreover, social connection between the pull-request submitter and project manager concerns the acceptance when the core team member is evaluating the pull-request~\cite{Tsay2014}.


Rejection of pull requests can increase when technical problems are not properly solving and if the number of forks increase too~\cite{Rahman2016}.
Other most important rejection factors are inexperience with pull requests; the complexity of contributions; the locality of the artifacts modified; and the project's policy contribution~\cite{Soares2015}.
From the integrator’s perspective, social challenges that needed to be addressed, for example, how to motivate contributors to keep working on the project and how to explain the reasons of rejection without discouraging them. From the contributor’s perspective, they found that it is important to reduce response time, maintain awareness, and improve communication~\cite{Gousios2014}.

\subsection{Software Quality of Pull Requests}
To the best of our knowledge, only a few studies have focused on the quality aspect of pull request acceptance ~\cite{Gousios2014}, ~\cite{Gousios2015}, ~\cite{Kononenko2018}. 

Gousios et al.~\cite{Gousios2014} investigated the pull-based development process focusing on the factors that affect the efficiency of the process and contribute to the acceptance of a pull request, and the related acceptance time. They analyzed the GHTorrent corpus and another 291 projects. The results showed that the number of pull requests increases over time. However, the proportion of repositories using them is relatively stable. They also identified common driving factors that affect the lifetime of pull requests and the merging process. Based on their study, code reviews did not seem to increase the probability of acceptance,  since 84\% of the reviewed pull requests were merged. 

Gousios et al.~\cite{Gousios2015} also conducted a survey aimed at characterizing the key factors considered in the decision-making process of pull request acceptance. Quality was revealed as one of the top priorities for developers. The most important acceptance factors they identified are: targeted area importance, test cases, and code quality. However, the respondents specified quality differently from their respective perception, as \textit{conformance}, \textit{good available documentation}, and \textit{contributor reputation}.

Kononenko et al.~\cite{Kononenko2018} investigated the pull request acceptance process in a commercial project addressing the quality of pull request reviews from the point of view of developers' perception. They applied data mining techniques on the project’s GitHub repository in order to understand the merge nature and then conducted a manual inspection of the pull requests.
They also investigated the factors that influence the merge time and outcome of pull requests such as pull request size and the number of people involved in the discussion of each pull request. Developers' experience and affiliation were two significant factors in both models.
Moreover, they report that developers generally associate the quality of a pull request  with the quality of its description, its complexity, and its revertability. However, they did not evaluate the reason for a pull request being rejected. 
These studies investigated the software quality of pull requests focusing on the trustworthiness of developers' experience and affiliation~\cite{Kononenko2018}. Moreover, these studies did not measure the quality of pull requests against a set of rules, but based on their acceptance rate and developers' perception. 
Our work complements these works by analyzing the code quality of pull requests in popular open-source projects and how the quality, specifically issues in the source code, affect the chance of a pull request being accepted when it is reviewed by a project maintainer. We measured code quality against a set of rules provided by PMD, one of the most frequently used open-source software tools for analyzing source code.


\section{Case Study Design}
\label{CS}
We designed our empirical study as a case study based on the guidelines defined by Runeson and H\"{o}st~\cite{Runeson2009}.
In this Section, we describe the case study design, including the goal and the research questions, the study context, the data collection, and the data analysis procedure. 

\subsection{Goal and Research Questions}
\label{RQ}
The goal of this work is to investigate the role of code quality in pull request acceptance. 

Accordingly, to meet our expectations, we formulated the goal as follows, using the Goal/Question/Metric (GQM) template~\cite{Basili1994}: \\

\begin{tabular}
{@{}p{1.5cm}p{7cm}@{}}
\textit{Purpose} & Analyze \\
\textit{Object} &  the acceptance of pull requests\\
\textit{Quality} & with respect to their code quality \\
\textit{Viewpoint} & from the point of view of developers \\ 
\textit{Context} & in the context of Java projects\\
 & \\
\end{tabular}

Based on the defined goal, we derived the following Research Questions (\textbf{RQs}): \\

\begin{tabular}
{@{}p{0.5cm}p{11cm}@{}}
\textbf{RQ1} & What is the distribution of TD items violated by the pull requests in the analyzed software systems? \\
\textbf{RQ2} & Does code quality affect pull request acceptance?\\
\textbf{RQ3} & Does code quality affect pull request acceptance considering different types and levels of severity of TD items?\\
 & \\
\end{tabular}

\textbf{RQ1} aims at assessing the distribution TD items violated by pull requests in the analyzed software systems. We also took into account the distribution of TD items with respect to their priority level as assigned by PMD (P1-P5). These results will also help us to better understand the context of our study. 

\textbf{RQ2} aims at finding out whether the project maintainers in open-source Java projects consider quality issues in the pull request source code when they are reviewing it. If code quality issues affect the acceptance of pull requests, the question is what kind of TD items errors generally lead to the rejection of a pull request.

\textbf{RQ3} aims at finding out if a severe code quality issue is more likely to result in the project maintainer rejecting the pull request. This will allow us to see whether project maintainers should pay more attention to specific issues in the code and make code reviews more efficient.

\subsection{Context}
\label{Context}
The projects for this study were selected using "criterion sampling"~\cite{Patton2002}. The criteria for selecting projects were as follows: 

\begin{itemize}
\item Uses Java as its primary programming language
\item Older than two years
\item Had active development in last year
\item Code is hosted on GitHub
\item Uses pull requests as a means of contributing to the code base
\item Has more than 100 closed pull requests
\end{itemize}

Moreover, we tried to maximize diversity and representativeness considering a comparable number of projects with respect to project age, size, and domain, as recommended by Nagappan et al.~\cite{Nagappan2013}.

We selected 28 projects according to these criteria. The majority, 22 projects, were selected from the Apache Software Foundation repository\footnote{http://apache.org}. The repository proved to be an excellent source of projects that meet the criteria described above. 
This repository includes some of the most widely used software solutions, considered industrial and mature, due to the strict review and inclusion process required by the ASF. Moreover, the included projects have to keep on reviewing their code and follow a strict quality process\footnote{https://incubator.apache.org/policy/process.html}.

The remaining six projects were selected with the help of the Trending Java repositories list that GitHub provides\footnote{https://github.com/trending/java}. GitHub provides a valuable source of data for the study of code reviews~\cite{Kalliamvakou2016}. In the selection, we manually selected popular Java projects using the criteria mentioned before.

In Table~\ref{table:projects}, we report the list of the 28 projects that were analyzed along with the number of pull requests (''\textit{\#PR}''), the time frame of the analysis, and the size of each project (\textit{''\#LOC'')}.
 
\begin{table}
\caption {Selected projects} \label{table:projects} 
\begin{center}
\footnotesize
 \begin{tabular}{p{4.8cm}|p{1cm}|p{2.5cm}|p{1.6cm}} 
 \hline
\textbf{Project Owner/Name} & \textbf{\#PR} & \textbf{Time Frame} &\textbf{\#LOC}\\ \hline

 apache/any23 & 129 & 2013/12-2018/11 & 78.35 \\
 \hline
 apache/dubbo & 1,27 & 2012/02-2019/01 & 133.63 \\
 \hline
 apache/calcite & 873 & 2014/07-2018/12 & 337.43\\
 \hline
 apache/cassandra & 182 & 2018/10-2011/09 & 411.24 \\
 \hline
 apache/cxf & 455 & 2014/03-2018/12 & 807.51\\
 \hline
 apache/flume & 180 & 2012/10-2018/12  & 103.70\\
 \hline
 apache/groovy & 833 & 2015/10-2019/01  & 396.43\\
 \hline
 apache/guacamole-client & 331 & 2016/03-2018/12 & 65.92\\
 \hline
 apache/helix & 284 & 2014/08-2018/11 & 191.83 \\
 \hline
 apache/incubator-heron & 2,19 & 2015/12-2019/01 & 207.36 \\
 \hline
 hibernate/hibernate-orm & 2,57 & 2010/10-2019/01 & 797.30 \\
 \hline
 apache/kafka & 5,52 & 2013/01-2018/12 & 376.68 \\
 \hline
 apache/lucene-solr & 264 & 2016/01-2018/12 & 1.416.20 \\ 
 \hline
 apache/maven & 166 & 2013/03-2018/12 & 107.80\\ 
 \hline
 apache/metamodel & 198 & 2014/09-2018/12 & 64.80 \\
 \hline
 mockito/mockito & 726 & 2012/11-2019/01 & 57.40\\
 \hline
 apache/netbeans & 1,02 & 2017/09-2019/01 & 6.115.97 \\
 \hline
 netty/netty & 4,12 & 2010/12-2019/01 & 275.97 \\
 \hline
 apache/opennlp & 330 & 2016/04-2018/12 & 136.54 \\
 \hline
 apache/phoenix & 203 & 2014/07-2018/12 & 366.58 \\
 \hline
 apache/samza & 1,47 & 2014/10-2018/10 & 129.28 \\
 \hline
 spring-projects/spring-framework & 1,85 & 2011/09-2019/01 & 717.96   \\
 \hline
 spring-projects/spring-boot & 3,07 & 2013/06-2019/01 & 348.09 \\
 \hline
 apache/storm & 2,86 & 2013/12-2018/12 & 359.90 \\
 \hline
 apache/tajo & 1,020 & 2014/03-2018/07 & 264.79 \\
 \hline
 apache/vxquery & 169 & 2015/04-2017/08 & 264.79 \\
 \hline
 apache/zeppelin & 3,19 & 2015/03-2018/12 & 218.95\\
 \hline
 openzipkin/zipkin & 1,47 & 2012/06-2019/01 & 121.50 \\
 \hline
 \textbf{Total} & \textbf{36,34} & \textbf{14.683.97} \\
 \hline
\end{tabular}
\end{center}
\end{table}

\subsection{Data Collection}
\label{DataCollection}

We first extracted all pull requests from each of the selected projects using the GitHub REST API v3 \footnote{https://developer.github.com/v3/}.

For each pull request, we fetched the code from the pull request's branch and analyzed the code using PMD. The default Java rule set for PMD was used for the static analysis. We filtered the TD items added in the main branch to only include items introduced in the pull request. The filtering was done with the aid of a diff-file provided by GitHub API and compared the pull request branch against the master branch.

We identified whether a pull request was accepted or not by checking whether the pull request had been marked as merged into the master branch or whether the pull request had been closed by an event that committed the changes to the master branch. Other ways of handling pull requests within a project were not considered.

\subsection{Data Analysis}
\label{DataAnalysis}
The result of the data collection process was a csv file reporting the dependent variable (pull request accepted or not) and the independent variables (number of TD items introduced in each pull request). Table~\ref{table:datasample} provides an example of the data structure we adopted in the remainder of this work. 

\begin{table}[ht]
\caption {Example of data structure used for the analysis} \label{table:datasample} 
\footnotesize
\centering
\begin{tabular}{l|l|l|l|l|l}
\hline
\multicolumn{2}{l|}{} & \textbf{Dependent Variable} & \multicolumn{3}{l}{\textbf{Independent Variables}} \\ \hline
\textbf{Project ID}   & \textbf{PR ID }  &\textbf{Accepted PR}        & \textbf{Rule1}         & ...        & \textbf{Rule n }       \\ \hline
Cassandra    & ahkji   & 1                  & 0             &            & 3             \\ \hline
Cassandra    & avfjo   & 0                  & 0             &            & 2             \\ \hline
\end{tabular}
\end{table}

For \textbf{RQ1}, we first calculated the total number of pull requests and the number of TD items present in each project. Moreover, we calculated the number of accepted and rejected pull requests. For each TD item, we calculated the number of occurrences, the number of pull requests, and the number of projects where it was found. Moreover, we calculated descriptive statistics (average, maximum, minimum, and standard deviation) for each TD item.

In order to understand if TD items affect pull request acceptance (\textbf{RQ2}), we first determined whether there is a significant difference between the expected frequencies and the observed frequencies in one or more categories. First, we computed the ${\chi}^2$ test. 
Then, we selected eight Machine Learning techniques and compared their accuracy. 
To overcome to the limitation of the different techniques, we selected and compared eight of them. The description of the different techniques, and the rationale adopted to select each of them is reported in Section~\ref{Background}.

${\chi}^2$ test could be enough to answer our RQs. However, in order to support possible follow-up of the work, considering other factors such as LOC as independent variable, Machine Learning techniques can provide much more accuracy results. 

We examined whether considering the priority value of an issue affects the accuracy metrics of the prediction models (\textbf{RQ3}). We used the same techniques as before but grouped all the TD items in each project into groups according to their priorities. The analysis was run separately for each project and each priority level (28 projects * 5 priority level groups) and the results were compared to the ones we obtained for RQ2. To further analyze the effect of issue priority, we combined the TD items of each priority level into one data set and created models based on all available items with one priority.

Once a model was trained, we confirmed that the predictions about pull request acceptance made by the model were accurate (\textbf{Accuracy Comparison}). To determine the accuracy of a model, 5-fold cross-validation was used. The data set was randomly split into five parts. A model was trained five times, each time using four parts for training and the remaining part for testing the model. We calculated accuracy measures (Precision, Recall, Matthews Correlation Coefficient, and F-Measure) for each model (see Table~\ref{table:metrics}) and then combined the accuracy metrics from each fold to produce an estimate of how well the model would perform.

We started by calculating the commonly used metrics, including F-measure, precision, recall, and the harmonic average of the latter two. Precision and recall are metrics that focus on the true positives produced by the model. Powers \cite{accuracy} argues that these metrics can be biased and suggests that a contingency matrix should be used to calculate additional metrics to help understand how negative predictions affect the accuracy of the constructed model. Using the contingency matrix, we calculated the model's Matthew Correlation Coefficient (MCC), which suggests as the best way to reduce the information provided by the matrix into a single probability describing the model's accuracy~\cite{accuracy}.


\newcolumntype{M}[1]{>{\centering\arraybackslash}m{#1}}
\newcolumntype{N}{@{}m{0pt}@{}}

\begin{table}
\caption {Accuracy measures} 
\label{table:metrics} 
\begin{center}
\footnotesize
\begin{tabular}{l|M{6.2cm}N}
\textbf{Accuracy Measure} & \textbf{Formula} & \\ 
\hline
\textbf{Precision} & \(\frac{TP}{FP + TP}\) &\\ [8pt]
\hline
\textbf{Recall} & \(\frac{TP}{FN + TP}\) &\\[8pt]
\hline
\textbf{MCC} & \(\frac{TP * TN - FP * FN}{\sqrt{(FP + TP)(FN - TP)(FP + TN)(FN + TN)}}\) &\\[8pt]
\hline
\textbf{F-measure} & \(2*\frac{precision * recall}{precision + recall}\) &\\[8pt]
\hline
\end{tabular}

\scriptsize{TP: True Positive; TN: True Negative; FP: False Positive; FN: False Negative}
\end{center}

\end{table}

For each classifier to easily gauge the overall accuracy of the machine learning algorithm in a model~\cite{BRADLEY19971145}, we calculated the Area Under The Receiver Operating Characteristic (AUC). For the AUC measurement, we calculated Receiver Operating Characteristics (ROC) and used these to find out the AUC ratio of the classifier, which is the probability of the classifier ranking a randomly chosen positive higher than a randomly chosen negative one. 

\subsection{Replicability}
\label{Replicability}
In order to allow our study to be replicated, we have published the complete raw data in the replication package\footnote{https://figshare.com/s/d47b6f238b5c92430dd7}.

\section{Results}
\label{Results}
\subsection*{RQ1.  What is the distribution of TD items violated by the pull requests in the analyzed software systems?}
For this study, we analyzed 36,344 pull requests violating 253 TD items and contained more than 4.7 million times (Table~\ref{table:PR-TDDiffuseness}) in the 28 analyzed projects. We found that 19,293 pull requests (53.08\%) were accepted and 17,051 pull requests (46.92\%) were rejected. Eleven projects contained the vast majority of the pull requests (80\%) and TD items (74\%). The distribution of the TD items differs greatly among the pull requests. For example, the projects \textit{Cassandra} and \textit{Phoenix} contain a relatively large number of TD items compared to the number of pull requests, while \textit{Groovy}, \textit{Guacamole}, and \textit{Maven} have a relatively small number of TD items. 

Taking into account the priority level of each rule, the vast majority of TD items (77.86\%) are classified with priority level 3, while the remaining ones (22.14\%) are equally distributed among levels 1, 2, and 4. None of the projects we analyzed had any issues rated as priority level 5.

Table~\ref{tab:Diffuseness} reports the number of TD items (''\textit{\#TD item}'') and their number of occurrences (''\textit{\#occurrences}'') grouped by priority level (''\textit{Priority}''). 

Looking at the TD items that could play a role in pull request acceptance or rejection, 243 of the 253 TD items (96\%) are present in both cases, while the remaining 10 are found only in cases of rejection (Table~\ref{tab:Diffuseness}).

Focusing on TD items that have with a ''double role'', we analyzed the distribution in each case. We discovered that 88 TD items have a diffusion rate of more than 60\% in the case of acceptance and 127 have a diffusion rate of more than 60\% in the case of rejection. The remaining 38 are equally distributed. 

Table~\ref{tab:Descriptive} and  Table~\ref{tab:Descriptive2} present preliminary information related to the twenty most recurrent TD items. We report descriptive statistics by means of Average (''\textit{Avg.}''), Maximum (''\textit{Max}''), Minimum (''\textit{Min}''), and Standard Deviation (''\textit{Std. dev.}''). Moreover, we include the priority of each TD item (''\textit{Priority}''), the sum of issue rows of that rule type found in the issues master table (''\textit{\# Total occurrences}''), and the number of projects in which the specific TD item has been violated (\textit{''\#Project''}). 

The complete list is available in the replication package (Section~\ref{Replicability}).
\begin{table} [H]
\caption {Distribution of pull requests (PR) and technical debt items (TD items) in the selected projects - (RQ1)} \label{table:PR-TDDiffuseness} 
\begin{center}
\footnotesize
 \begin{tabular}{p{4cm}|p{1.2cm}|p{2cm}|p{1.3cm}|p{1.3cm}} 
 \hline
\textbf{Project Name} & \textbf{\#PR} & \textbf{\#TD Items} & \textbf{\% Acc.} & \textbf{\% Rej.} \\ \hline
 apache/any23 & 129  & 11,573 & 90.70 & 9.30 \\
 \hline
 apache/dubbo & 1,270 &  169,751 & 52.28 & 47.72 \\
 \hline
 apache/calcite & 873 & 104,533 & 79.50 & 20.50\\
 \hline
 apache/cassandra & 182  & 153,621 & 19.78 & 80.22 \\
 \hline
 apache/cxf & 455  & 62,564 & 75.82 & 24.18 \\
 \hline
 apache/flume & 180 & 67,880 & 60.00 & 40.00 \\
 \hline
 apache/groovy & 833  & 25,801 & 81.39 & 18.61 \\
 \hline
 apache/guacamole-client & 331  & 6,226 & 92.15 & 7.85\\
 \hline
 apache/helix & 284 & 58,586 & 90.85 & 9.15 \\
 \hline
 apache/incubator-heron & 2,191  & 138,706 & 90.32 & 9.68 \\
 \hline
 hibrenate/hibernate-orm & 2,573  & 490,905 & 16.27 & 83.73 \\
 \hline
 apache/kafka & 5,522 &  507,423 & 73.51 & 26.49 \\
 \hline
 apache/lucene-solr & 264 & 72,782 & 28.41 & 71.59 \\ 
 \hline
 apache/maven & 166 &  4,445 & 32.53 & 67.47 \\ 
 \hline
 apache/metamodel & 198 & 25,549 & 78.28 & 21.72 \\
 \hline
 mockito/mockito & 726 &  57,345 & 77.41 & 22.59 \\
 \hline
 apache/netbeans & 1,026 &  52,817 & 83.14 & 16.86 \\
 \hline
 netty/netty & 4,129 & 597,183 & 15.84 & 84.16 \\
 \hline
 apache/opennlp & 330 &  21,921 & 82.73 & 17.27 \\
 \hline
 apache/phoenix & 203 & 214,997 & 9.85 & 90.15 \\
 \hline
 apache/samza & 1,475 & 96,915 & 69.52 & 30.48 \\
 \hline
 spring-projects/spring-framework & 1,850 & 487,197 & 15.68 & 84.32   \\
 \hline
 spring-projects/spring-boot & 3,076  & 156,455 & 8.03 & 91.97 \\
 \hline
 apache/storm & 2,863 & 379,583 & 77.96 & 22.04 \\
 \hline
 apache/tajo & 1,020 & 232,374 & 67.94 & 32.06 \\
 \hline
 apache/vxquery & 169 & 19,033 & 30.77 & 69.23 \\
 \hline
 apache/zeppelin & 3,194 &  408,444 & 56.92 & 43.08 \\
 \hline
 openzipkin/zipkin & 1,474 &  78,537 & 73.00 & 27.00 \\
 \hline
 \textbf{Total} & \textbf{36,344} &  \textbf{4,703,146} & \textbf{19,293} & \textbf{17,051} \\
 \hline
\end{tabular}
\end{center}
\end{table}

\begin{table}[H]
\vspace{-1mm}
\centering 
\caption{Distribution of TD items in pull requests - (RQ1)}
\footnotesize
\label{tab:Diffuseness}
\begin{tabular}
{p{1.3cm}|p{2cm}|p{2.4cm}|p{1.8cm}|p{1.82cm}}
\hline 
\textbf{Priority} &	\textbf{\#TD Items} &	\textbf{\#occurrences} &	\textbf{\% PR Acc.}&	\textbf{\% PR Rej.}	\\	\hline
All		&	253	&	4,703,146 & 96.05 & 100.00\\	\hline \hline 
4		&	18	&	85,688 & 77.78 & 100.00\\	\hline
3		&	197	&	4,488,326 & 96.95 & 100.00\\	\hline
2		&	22	&	37,492 & 95.45 & 95.45\\	\hline
1		&	16  &	91,640	& 100.00 & 100.00\\	\hline
\end{tabular}
\end{table}

\vspace*{5mm}
\hspace*{-5mm}
\begin{tikzpicture}
\node [mybox] (box){%
\centering
\begin{minipage}{.95\textwidth}
\textbf{Summary of RQ1}\\
Among the 36,344 analyzed pull requests, we discovered 253 different type of TD items (PMD Rules) violated more that 4.7 million times. Nearly half of the pull requests had been accepted and the other half had been rejected. 243 of the 253 TD items were found to be present in both cases.
The vast majority of these TD items (197) have priority level 3. 
\end{minipage}
};
\end{tikzpicture}%

\subsection*{RQ2. Does code quality affect pull request acceptance?}
To answer this question, we trained machine learning models for each project using all possible pull requests at the time and using all the different classifiers introduced in Section~\ref{Background}. A pull request was used if it contained Java that could be analyzed with PMD. There are some projects in this study that are multilingual, so filtering of the analyzable pull requests was done out of necessity.

Once we had all the models trained, we tested them and calculated the accuracy measures described in Table~\ref{table:metrics} for each model. We then averaged each of the metrics from the classifiers for the different techniques. The results are presented in Table~\ref{tab:Accuracy}. The averaging provided us with an estimate of how accurately we could predict whether maintainers accepted the pull request based on the number of different TD items it has.
The results of this analysis are presented in Table~\ref{tab:results}. For reasons of space, we report only the most frequent 20 TD items. The table also contains the number of distinct PMD rules that the issues of the project contained. The rule count can be interpreted as the number of different types of issues found.


\begin{table}[H]
\centering
\caption{Model reliability - (RQ2)}
\label{tab:Accuracy}
\footnotesize
\begin{tabular}{p{1.6cm}|p{0.8cm}|p{0.9cm}|p{1cm}|p{0.9cm}|p{0.8cm}|p{0.9cm}|p{0.9cm}|p{1cm}}
\hline
& \multicolumn{8}{c}{\textbf {Average between 5-fold validation models}} \\ \cline{2-9}
\textbf{Accuracy Measure} & \textbf{L. R.}& \textbf{D. T.}& \textbf{Bagg.}& \textbf{R. F.}& \textbf{E. T.} & \textbf{A. B.}  & \textbf{G. B.} & \textbf{XG.B.} \\
\hline
AUC & 50.91 & 50.12 & 49.83 & 50.75 & 50.54 & 51.30 & 50.64 & 50.92 \\ \hline
Precision & 49.53 & 48.40 & 48.56  & 49.33 & 49.20  & 48.74 & 49.30  & 49.20  \\ \hline
RECALL & 62.46  & 47.45  & 47.74 & 48.07 & 47.74 & 51.82 & 41.80 & 41.91  \\ \hline
MCC & 0.02  & -0.00  & 0.00  & 0.01  & 0.01  & 0.00  & 0.00  & -0.00  \\ \hline
F-Measure & 0.55  & 0.47  & 0.47  & 0.48 & 0.48  & 0.49  & 0.44  & 0.44 \\ \hline
\end{tabular}
\end{table}

\begin{table}
\vspace{-5mm}
\centering 
\footnotesize
\caption{Descriptive statistics (the 15 most recurrent TD items) - Priority, number of occurrences (\#occur.), number of Pull Requests (\#PR) and number of projects (\#prj.)- (RQ1)}
\label{tab:Descriptive}
\begin{tabular}
{p{5.7cm}|p{1.2cm}|p{1.5cm}|p{1cm}|p{0.8cm}} 
\hline 
\textbf{TD Item}& \textbf{Priority} & \textbf{\#occur.} & \textbf{\#PR} & \textbf{\#prj.} \\ \hline
LawOfDemeter	&	4	&	1,089,110	&	15,809	&	28		\\	\hline
MethodArgumentCouldBeFinal	&	4	&	627,688	&	12,822	&	28	\\	\hline
CommentRequired	&	4	&	584,889	&	15,345	&	28		\\	\hline
LocalVariableCouldBeFinal	&	4	&	578,760	&	14,920	&	28		\\	\hline
CommentSize	&	4	&	253,447	&	11,026	&	28		\\	\hline
JUnitAssertionsShouldIncludeMessage	&	4	&	196,619	&	6,738	&	26		\\	\hline
BeanMembersShouldSerialize	&	4	&	139,793	&	8,865	&	28		\\	\hline
LongVariable	&	4	&	122,881	&	8,805	&	28		\\	\hline
ShortVariable	&	4	&	112,333	&	7,421	&	28	\\	\hline
OnlyOneReturn	&	4	&	92,166	&	7,111	&	28	\\	\hline
CommentDefaultAccessModifier	&	4	&	58,684	&	5,252	&	28		\\	\hline
DefaultPackage	&	4	&	42,396	&	4,201	&	28	\\	\hline
ControlStatementBraces	&	4	&	39,910	&	2,689	&	27		\\	\hline
JUnitTestContainsTooManyAsserts	&	4	&	3,6022	&	4,954	&	26		\\	\hline
AtLeastOneConstructor	&	4	&	29,516	&	5,561	&	28	\\	\hline
\end{tabular}
\end{table}

\begin{table}
\vspace{-5mm}
\centering 
\footnotesize
\caption{Descriptive statistics (the 15 most recurrent TD items) - Average (Avg.), Maximum (Max), Minimum (Min) and Standard Deviation (std. dev.) - (RQ1)}
\label{tab:Descriptive2}
\begin{tabular}
{p{5.7cm}|p{1.6cm}|p{1.2cm}|p{0.8cm}|p{1.8cm}} 
\hline 
\textbf{TD Item}& \textbf{Avg} & \textbf{Max} & \textbf{Min} & \textbf{Std. dev.} \\ \hline
LawOfDemeter	&	38,896.785	&	140,870	&	767	&	40,680.62855	\\	\hline
MethodArgumentCouldBeFinal		&	22,417.428	&	105,544	&	224	&	25,936.63552	\\	\hline
CommentRequired		&	20,888.892	&	66,798	&	39	&	21,979.94058	\\	\hline
LocalVariableCouldBeFinal	&	20,670	&	67394	&	547	&	20,461.61422	\\	\hline
CommentSize		&	9,051.678	&	57,074	&	313	&	13,818.66674	\\	\hline
JUnitAssertionsShouldIncludeMessage	&	7,562.269	&	38,557	&	58	&	10822.38435	\\	\hline
BeanMembersShouldSerialize		&	4,992.607	&	22,738	&	71	&	5,597.458969	\\	\hline
LongVariable		&	4,388.607	&	19,958	&	204	&	5,096.238761	\\	\hline
ShortVariable		&	4,011.892	&	21,900	&	26	&	5,240.066577	\\	\hline
OnlyOneReturn		&	3,291.642	&	14,163	&	42	&	3,950.4539	\\	\hline
CommentDefaultAccessModifier		&	2,095.857	&	12,535	&	6	&	2,605.756401	\\	\hline
DefaultPackage	&	1,514.142	&	9,212	&	2	&	1,890.76723	\\	\hline
ControlStatementBraces		&	1,478.148	&	11,130	&	1	&	2,534.299929	\\	\hline
JUnitTestContainsTooManyAsserts		&	1,385.461	&	7,888	&	7	&	1,986.528192	\\	\hline
AtLeastOneConstructor	&	1,054.142	&	6,514	&	21	&	1,423.124177	\\	\hline
\end{tabular}
\end{table}


\begin{landscape}
\begin{table*}
\centering 
\footnotesize
\caption{Summary of the quality rules related to pull request acceptance - (RQ2 and RQ3)}
\label{tab:results}
\footnotesize
\begin{tabular}{p{5.4cm}|p{0.9cm}|p{0.8cm}|p{1.3cm}|p{0.8cm}|p{0.8cm}|p{0.8cm}|p{0.8cm}|p{0.8cm}|p{0.8cm}|p{0.8cm}|p{1cm}}
\hline 
\multirow{2}{*}{\textbf{Rule ID}} & \multirow{2}{*}{\textbf{Prior.}} & \multirow{2}{*}{\textbf{\#prj.}} & \multirow{2}{*}{\textbf{\#occur.}} & \multicolumn{8}{c}{\textbf{Importance (\%)}} \\  \cline{5-12}
& & & &  \textbf{A.B.} & \textbf{Bagg.} & \textbf{D.T.} & \textbf{E.T.}&  \textbf{G.B.} &  \textbf{L.R.} & \textbf{R.F.} &\textbf{XG.B.}\\ \hline 
LawOfDemeter &  4 &  28 &  1089110 &  0.12  &  -0.51  &  0.77  &  -0.74  &  -0.29  &  -0.09  &  -0.66  &  0.02 \\ \hline
MethodArgumentCouldBeFinal &  4 &  28 &  627688 &  -0.31  &  0.38  &  0.14  &  0.03  &  -0.71  &  -0.25  &  0.24  &  0.07 \\ \hline
CommentRequired &  4 &  28 &  584889 &  -0.25  &  -0.11  &  0.07  &  -0.30  &  -0.47  &  -0.17  &  0.58  &  -0.31 \\ \hline
LocalVariableCouldBeFinal &  4 &  28 &  578760 &  -0.13  &  -0.20  &  0.55  &  0.28  &  0.08  &  -0.05  &  0.61  &  -0.05 \\ \hline
CommentSize &  4 &  28 &  253447 &  -0.24  &  -0.15  &  0.49  &  -0.08  &  -0.17  &  -0.05  &  -0.10  &  0.05 \\ \hline
JUnitAssertionsShouldIncludeMessage &  4 &  26 &  196619 &  -0.41  &  -0.84  &  0.22  &  -0.28  &  -0.19  &  -0.10  &  -0.75  &  0.14 \\ \hline
BeanMembersShouldSerialize &  4 &  28 &  139793 &  -0.33  &  -0.09  &  -0.03  &  -0.38  &  -0.37  &  0.17  &  0.26  &  0.07 \\ \hline
LongVariable &  4 &  28 &  122881 &  0.08  &  -0.19  &  -0.02  &  -0.25  &  -0.28  &  0.08  &  0.24  &  0.02 \\ \hline
ShortVariable &  4 &  28 &  112333 &  -0.51  &  -0.24  &  0.09  &  -0.04  &  -0.04  &  0.07  &  -0.25  &  -0.54 \\ \hline
OnlyOneReturn &  4 &  28 &  92166 &  -0.69  &  -0.03  &  0.02  &  -0.25  &  -0.08  &  -0.06  &  0.06  &  -0.13 \\ \hline
CommentDefaultAccessModifier &  4 &  28 &  58684 &  -0.17  &  -0.07  &  0.30  &  -0.41  &  -0.25  &  0.23  &  0.18  &  -0.10 \\ \hline
DefaultPackage &  4 &  28 &  42396 &  -0.37  &  -0.05  &  0.20  &  -0.23  &  -0.93  &  0.10  &  -0.01  &  -0.54 \\ \hline
ControlStatementBraces &  4 &  27 &  39910 &  -0.89  &  0.09  &  0.58  &  0.29  &  -0.37  &  -0.03  &  0.08  &  0.25 \\ \hline
JUnitTestContainsTooManyAsserts &  4 &  26 &  36022 &  0.40  &  0.22  &  -0.25  &  -0.33  &  0.01  &  0.16  &  0.10  &  -0.17 \\ \hline
AtLeastOneConstructor &  4 &  28 &  29516 &  0.00  &  -0.29  &  -0.06  &  -0.18  &  -0.19  &  -0.07  &  0.15  &  -0.22 \\ \hline
UnnecessaryFullyQualifiedName &  4 &  27 &  27402 &  0.00  &  0.08  &  0.25  &  -0.05  &  0.00  &  0.00  &  0.26  &  -0.11 \\ \hline
AvoidDuplicateLiterals &  4 &  28 &  27224 &  -0.20  &  0.05  &  0.33  &  -0.28  &  0.12  &  0.20  &  0.09  &  0.07 \\ \hline
SignatureDeclareThrowsException &  4 &  27 &  26188 &  -0.18  &  -0.10  &  0.04  &  -0.13  &  -0.05  &  0.11  &  0.33  &  -0.17 \\ \hline
AvoidInstantiatingObjectsInLoops &  3 &  28 &  25344 &  -0.05  &  0.07  &  0.43  &  -0.14  &  -0.27  &  -0.13  &  0.52  &  -0.07 \\ \hline
FieldNamingConventions &  3 &  28 &  25062 &  0.09  &  0.00  &  0.16  &  -0.21  &  -0.10  &  -0.01  &  0.07  &  0.19 \\ \hline
\end{tabular}
\end{table*}
\end{landscape}

\begin{table}[H]
\centering
\caption{Contingency matrix}
\label{tab:ContingencyMatrix}
\footnotesize
\begin{tabular}{c|c|c}
\cline{2-3}
& \textbf{TD items} &  \textbf{No TD items} \\ \hline 
\textbf{PR accepted} & 10.563 & 8.558 \\\hline 
\textbf{PR rejected} & 11.228 & 5.528 \\ \hline 
\end{tabular}
\end{table}

\begin{figure}[H]
  \centering 
  \includegraphics[width=0.8\textwidth]{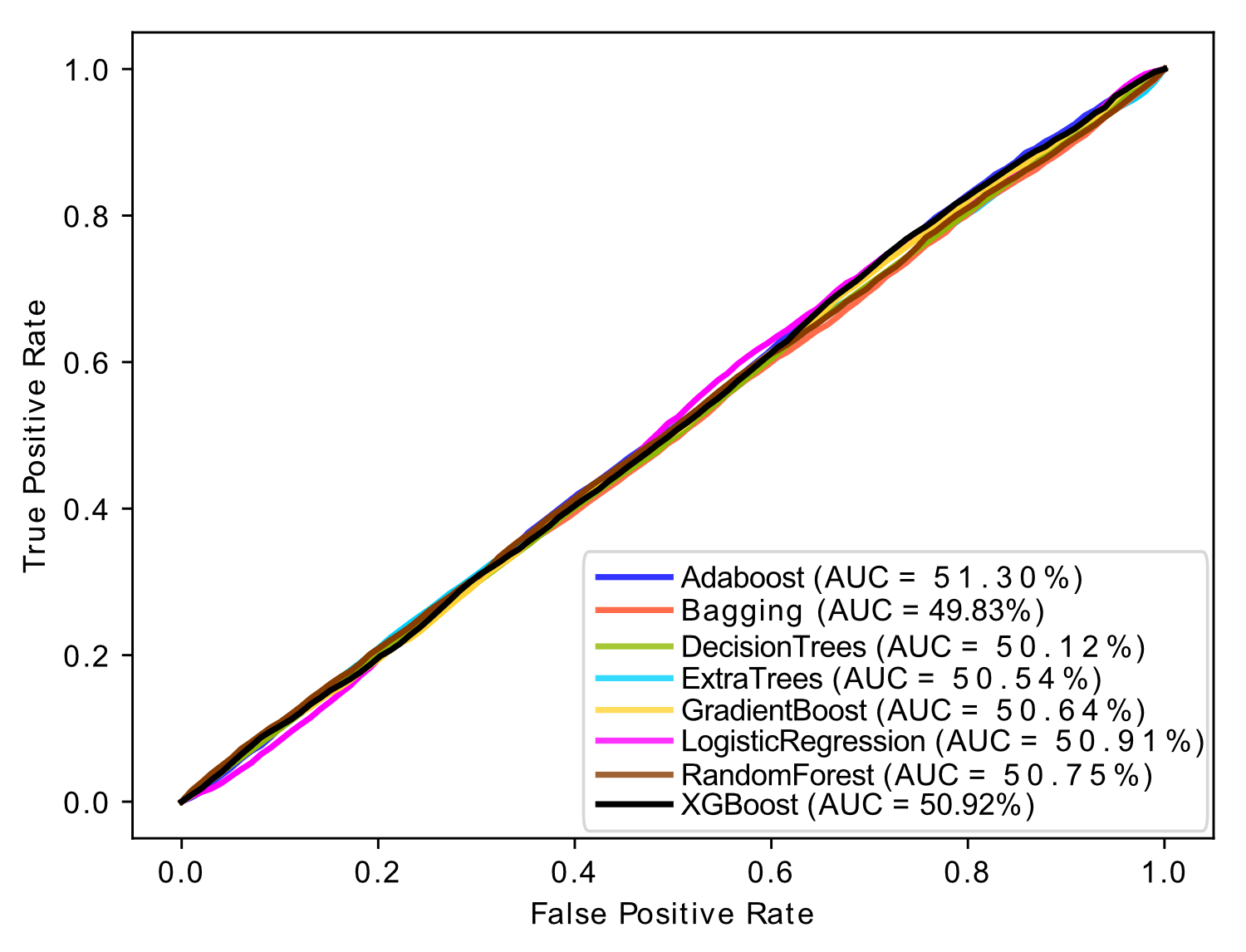}
  \caption{ROC Curves (average between 5-fold validation models) - (RQ2)}
  \label{fig:ROC}
\end{figure}

As depicted in Figure~\ref{fig:ROC}, almost all of the models' AUC for every method of prediction hovering around 50\%, overall code quality does not appear to be a factor in determining whether a pull request is accepted or rejected.

There were some projects that showed some moderate success, but these can be dismissed as outliers.

The results can suggest that perhaps Machine Learning could not be the most suitable techniques. However, also  ${\chi}^2$ test on the contingency matrix (0.12) (Table~\ref{tab:ContingencyMatrix}) confirms the above results that the presence of TD items does not affect pull request acceptance (which means that TD items and pull request acceptance are mutually independent). 


\subsection*{RQ3. Does code quality affect pull request acceptance considering different types and levels of severity of TD items?}

To answer this research question, we introduced PMD priority values assigned to each TD item. By taking these priorities into consideration, we grouped all issues by their priority value and trained the models using data composed of only issues of a certain priority level. 

Once we had run the training and tested the models with the data grouped by issue priority, we calculated the accuracy metrics mentioned above. These results enabled us to determine whether the prevalence of higher-priority issues affects the accuracy of the models. The affect on model accuracy or \textit{importance} is determined with the use of \textit{drop-column importance} -mechanism\footnote{https://explained.ai/rf-importance/}. After training our baseline model with P amount of features, we trained P amount of new models and compared each of the new models' tested accuracy against the baseline model. Should a feature affect the accuracy of the model, the model trained with that feature dropped from the dataset would have a lower accuracy score than the baseline model. The more the accuracy of the model drops with a feature removed, the more important that feature is to the model when classifying pull-requests as accepted or rejected. In table \ref{tab:results} we show the importance of the 20 most common quality rules when comparing the baseline model accuracy with a model that has the specific quality rule dropped from the feature set.

Grouping by different priority levels did not provide any improvement of the results in terms of accuracy. 

\vspace*{5mm}
\hspace*{-5mm}
\begin{tikzpicture}
\node [mybox] (box){%
\centering
\begin{minipage}{.95\textwidth}
\textbf{Summary of RQ2 and RQ3}\\
Looking at the results we obtained from the analysis using statistical and machine learning techniques (${\chi}^2$ 0.12 and AUC 50\% on average), code quality does not appear to influence pull request acceptance. 
\end{minipage}
};
\end{tikzpicture}%
\section{Discussion}
\label{Discussion}

In this Section, we will discuss the results obtained according to the RQs and present possible practical implications from our research. 

The analysis of the pull requests in 28 well-known Java projects shows that code quality, calculated by means of PMD rules, is not a driver for the acceptance or the rejection of pull requests. 
PMD recommends manual customization of the set of rules instead of using the out-of-the-box rule set and selecting the rules that developers should consider in order to maintain a certain level of quality. However, since we analyzed all the rules detected by PMD, no rule would be helpful and any customization would be useless in terms of being able to predict the software quality in code submitted to a pull request. 
The result cannot be generalized to all the open source and commercial projects, as we expect some project could enforce quality checks to accept pull requests. Some tools, such as SonarQube (one of the main PMD competitor), recently launched a new feature to allow developers to check the TD Issues before submitting the pull requests. Even if maintainers are not sensible to the quality of the code to be integrated in their projects, at least based on the rules detected by PMD, the adoption of pull request quality analysis tools such as SonarQube or the usage of PMD before submitting a pull request will increase the quality of their code, increasing the overall software maintainability and decreasing the fault proneness that could be increased from the injection of some TD items (see Table I).

The results complement those obtained by Soares et al. ~\cite{Soares2015} and Calefato et al. \cite{Calefato2017}, namely, that the reputation of the developer might be more important than the quality of the code developed. 
The main implication for practitioners, and especially for those maintaining open-source projects, is the realization that they should pay more attention to software quality. Pull requests are a very powerful instrument, which could provide great benefits if  they were used for code reviews as well.  
Researchers should also investigate whether other quality aspects might influence the acceptance of pull requests. 
\section{Threats to Validity}
\label{Threats}
In this Section, we will introduce the threats to validity and the different tactics we adopted to mitigate them, 

\textbf{Construct Validity}. This threat concerns the relationship between theory and observation due to possible measurement errors. 
Above all, we relied on PMD, one of the most used software quality analysis tool for Java. However, beside PMD is  largely used in industry, we did not find any evidence or empirical study assessing its detection accuracy. Therefore, we cannot exclude the presence of false positive and false negative in the detected TD items.
We extracted the code submitted in pull requests by means of the GitHub API\textsuperscript{10}. However, we identified whether a pull request was accepted or not by checking whether the pull request had been marked as merged into the master branch or whether the pull request had been closed by an event that committed the changes to the master branch. Other ways of handling pull requests within a project were not considered and, therefore, we are aware that there could be the limited possibility that some maintainer could have integrated the pull request code into their projects manually, without marking the pull request as accepted.


\textbf{Internal Validity}. This threat concerns internal factors related to the study that might have affected the results.
In order to evaluate the code quality of pull requests, we applied the rules provided by PMD, which is  one  of  the  most widely used static code analysis tools for Java on the market, also considering the different severity levels of each rule provided by PMD.  We are aware that the presence or the absence of a PMD issue cannot be the perfect predictor for software quality, and other rules or metrics detected by other tools could have brought to different results.  

\textbf{External Validity}. This threat concerns the generalizability of the results. We selected 28 projects. 21 of them were from the Apache Software Foundation, which incubates only certain systems that follow specific and strict quality rules. The remaining six projects were selected  with the help of the trending Java repositories  list provided by GitHub. In the selection, we preferred projects that are considered ready for production environments and are using pull requests as a way of taking in contributions.
Our case study was not based only on one application domain. This was avoided since we aimed to find general mathematical models for the prediction of the number of bugs in a system. Choosing only one domain or a very small number of application domains could have been an indication of the non-generality of our study, as only prediction models from the selected application domain would have been chosen. The selected projects stem from a very large set of application domains, ranging from external libraries, frameworks, and web utilities to large computational infrastructures. The application domain was not an important criterion for the selection of the projects to be analyzed, but at any rate we tried to balance the selection and pick systems from as many contexts as possible. However, we are aware that other projects could have enforced different quality standards, and could use different quality check before accepting pull requests. 
Furthermore, we are considering only open source projects, and we cannot speculate on industrial projects, as different companies could have different internal practices. Moreover, we also considered only Java projects. The replication of this work on different languages and different projects may bring to different results. 

\textbf{Conclusion Validity}. This threat concerns the relationship between the treatment and the outcome. In our case, this threat could be represented by the analysis method applied in our study. We reported the results considering descriptive statistics. Moreover, instead of using only Logistic Regression, we compared the prediction power of  different classifier to reduce the bias of the low prediction power that one single classifier could have. We do not exclude the possibility that other statistical or machine learning approaches such as Deep Learning or others might have yielded similar or even better accuracy than our modeling approach. However, considering the extremely low importance of each TD Issue and its statistical significance, we do not expect to find big differences applying other type of classifiers.

\section{Conclusion}
\label{Conclusion}
Previous works reported 84\% of pull requests to be accepted based on the trustworthiness of the developers~\cite{Gousios2015}\cite{Calefato2017}. However, pull requests are one of the most common code review mechanisms, and we believe that open-source maintainers are also considering the code quality when accepting or rejecting pull requests. 

In order to verify this statement, we analyzed the code quality of pull requests by means of PMD, one of the most widely used static code analysis tools, which can detect different types of quality flaws in the code (TD Issues), including design flaws, code smells, security vulnerability, potential bugs, and many other issues.  We considered PMD as it is able to detect a good number of TD Issues of different types that have been empirically considered harmful by several works.  Examples of these TD Issues are  God Class, High Cyclomatic Complexity, Large Class and Inappropriate Intimacy. 

We applied basic statistical techniques, but also eight machine learning classifiers to understand if it is possible to predict if a pull request could be accepted or not based on the presence of a set of TD Issue in the pull request code. 
Of the 36,344 pull requests we analyzed in 28 well-known Java projects, nearly half had been accepted and the other half rejected. 243 of the 253 TD items were present in each case.

Unexpectedly, the presence of TD items of all types in the pull request code, does not influence the acceptance or rejection of pull requests at all and therefore, \textbf{the quality of the code submitted in a pull request does not influence at all its acceptance}. The same results are verified in all the 28 projects independently. Moreover, also merging all the data as a single large data-set confirmed the results.

Our results complement the conclusions derived by  Gausios et al.~\cite{Gousios2015} and Calefato et al.~\cite{Calefato2017}, who report that the reputation of the developer submitting the pull request is one of the most important acceptance factors. 

As future work, we plan to investigate whether there are other types of qualities that might affect the acceptance of pull requests, considering TD Issues and metrics detected by other tool, analyzing different projects written in different languages. We also will also investigate how to raise awareness in the open-source community that code quality should also be considered when accepting pull requests. 

Moreover, we will understand the perceived harmfulness of developers about PMD rules, in order to qualitatively assess over these violations. Another important factor need to be consider is the developers' personality as possible influence on the acceptance of the pull request~\cite{CalefatoIST}.

\section*{References}
\bibliographystyle{model1-num-names}
\bibliography{references}

\end{document}